\ifdefined\isarxivversion
\documentclass[11pt]{article}

\usepackage[numbers]{natbib}
\usepackage{amsthm}
\usepackage{subfig}
\usepackage{wrapfig,epsfig}

\else

\documentclass[conference]{IEEEtran}
\IEEEoverridecommandlockouts

\def\BibTeX{{\rm B\kern-.05em{\sc i\kern-.025em b}\kern-.08em
    T\kern-.1667em\lower.7ex\hbox{E}\kern-.125emX}}
    
\fi

\usepackage{amsthm}
\usepackage{amsmath}
\usepackage{amssymb}
\usepackage{algorithm}
\usepackage{algpseudocode}
\usepackage{graphicx}
\usepackage{grffile}
\usepackage{url}
\usepackage{xcolor}
\usepackage{epstopdf}

\usepackage{bbm}
\usepackage{dsfont}
\usepackage{comment}

\allowdisplaybreaks

\ifdefined\isarxivversion

\let\C\relax
\usepackage{tikz}
\usepackage{hyperref}  
\hypersetup{colorlinks=true,citecolor=blue,linkcolor=blue} 
\usetikzlibrary{arrows}
\usepackage[margin=1in]{geometry}

\else
\usepackage{microtype}
\usepackage{hyperref}
\definecolor{mydarkblue}{rgb}{0,0.08,0.45}
\hypersetup{colorlinks=true, citecolor=mydarkblue,linkcolor=mydarkblue}

\fi

\newtheorem{theorem}{Theorem}[section]
\newtheorem{lemma}[theorem]{Lemma}
\newtheorem{definition}[theorem]{Definition}

\newtheorem{claim}[theorem]{Claim}

\newcommand{\wh}{\widehat}
\newcommand{\wt}{\widetilde}
\newcommand{\ov}{\overline}
\newcommand{\eps}{\epsilon}

\newcommand{\R}{\mathbb{R}}

\renewcommand{\i}{\mathbf{i}}
\renewcommand{\varepsilon}{\epsilon}

\renewcommand{\hat}{\wh}

\DeclareMathOperator*{\E}{{\mathbb{E}}}

\DeclareMathOperator*{\C}{\mathbb{C}}

\DeclareMathOperator{\supp}{supp}
\DeclareMathOperator{\poly}{poly}

\makeatletter
\newcommand*{\RN}[1]{\expandafter\@slowromancap\romannumeral #1@}
\makeatother

\DeclareMathOperator{\Err}{Err}

\usepackage{lineno}

\title{An $O(k\log n)$ Time Fourier Set Query Algorithm}

\begin{document}

\ifdefined\isarxivversion
\date{}

\author{
}

\else

\author{\IEEEauthorblockN{Yeqi Gao}
\IEEEauthorblockA{\textit{Department of CS} \\
\textit{University of Washington}\\
Seattle, USA \\
a916755226@gmail.com
}
\and
\IEEEauthorblockN{Zhao Song}
\IEEEauthorblockA{
\textit{Adobe Research}\\
\textit{Adobe}\\
San Jose, USA \\
zsong@adobe.com} 
\and 
\IEEEauthorblockN{Baocheng Sun}
\IEEEauthorblockA{\textit{Department of CS \& Applied Math } \\
\textit{Weizmann Institute of Science}\\
Rehovot, Israel \\
woafrnraetns@gmail.com}
}

\maketitle

\fi

\ifdefined\isarxivversion
\begin{titlepage}
  \maketitle
  \begin{abstract}
Fourier transformation is an extensively studied problem in many research fields. It has many applications in machine learning, signal processing, compressed sensing, and so on. In many real-world applications, approximated Fourier transformation is sufficient and we only need to do the Fourier transform on a subset of coordinates. 
Given a vector $x \in \mathbb{C}^{n}$, an approximation parameter $\epsilon$ and a query set $S \subset [n]$ of size $k$, we propose an algorithm to compute an approximate Fourier transform result $x'$ which uses $O(\epsilon^{-1} k \log(n/\delta))$ Fourier measurements, runs in $O(\epsilon^{-1} k \log(n/\delta))$ time and outputs a vector $x'$ such that $\| ( x' - \widehat{x} )_S  \|_2^2 \leq \epsilon \| \widehat{x}_{\bar{S}} \|_2^2 + \delta \| \widehat{x} \|_1^2 $ holds with probability of at least $9/10$.

  \end{abstract}
  \thispagestyle{empty}
\end{titlepage}

\newpage

\else
\begin{abstract}
Fourier transformation is an extensively studied problem in many research fields. It has many applications in machine learning, signal processing, compressed sensing, and so on. In many real-world applications, approximated Fourier transformation is sufficient and we only need to do the Fourier transform on a subset of coordinates. 
Given a vector $x \in \mathbb{C}^{n}$, an approximation parameter $\epsilon$ and a query set $S \subset [n]$ of size $k$, we propose an algorithm to compute an approximate Fourier transform result $x'$ which uses $O(\epsilon^{-1} k \log(n/\delta))$ Fourier measurements, runs in $O(\epsilon^{-1} k \log(n/\delta))$ time and outputs a vector $x'$ such that $\| ( x' - \widehat{x} )_S  \|_2^2 \leq \epsilon \| \widehat{x}_{\bar{S}} \|_2^2 + \delta \| \widehat{x} \|_1^2 $ holds with probability of at least $9/10$.

\end{abstract}

\fi

\section{Introduction}

 Fourier transform is ubiquitous in image and audio processing, telecommunications and so on. The time complexity of classical Fast Fourier Transform (FFT) algorithm proposed by Cooley and Turkey~\cite{ct65} is $O(n \log n)$. Optics imaging~\cite{v11, g17}, magnetic resonance image (MRI)~\cite{assn08} and the physics~\cite{r89} are benefits from this algorithm. 
 The algorithm proposed by Cooley and Turkey~\cite{ct65} takes $O(n)$ samples to compute the Fourier transformation result.

The number of samples taken is an important factor. For example, it influences the amount of ionizing radiation that a patient is exposed to during CT scans. The amount of time a patient spends within the scanner can also be reduced by taking fewer samples. Thus, we consider the Fourier Transform problems in two computational aspects. 
Thus, two aspects of the Fourier Transform problems are taken into consideration by us. The first aspect is the reconstruction time which is the time of decoding the signal from the measurements. The second aspect is the sample complexity. Sample complexity is the number of noisy samples required by the algorithm. 
There is a long line of works optimizing the time and the sample complexity of Fourier Transform in the field of signal-processing and the field of TCS \cite{ct65,r89,assn08,v11,hikp12a,b15}. 

As a result, we can anticipate that algorithms that leverage sparsity assumptions about the input and outperform FFT in applications will be of significant practical utility. In general, the two most significant factors to optimize are the sample complexity and the time complexity of obtaining the Fourier Transform result.

In many real world applications, computing the approximate Fourier transformation results for a set of selective coordinates is sufficient, and we can leverage the approximation guarantee to accelerate the computation. The set query is originally proposed by \cite{p11}. The original definition doesn't have restriction on Fourier measurements. Then \cite{k17} generalizes the classical set query definition \cite{p11} into Fourier setting. 
In this paper we consider the set estimation based on Fourier measurement problem (defined by \cite{k17}) where given a vector $x \in \C^n$,  approximation parameters $\epsilon, \delta \in (0,1)$ and a query set $S \subset [n]$ and $|S| = k$, we want to compute an approximate Fourier transform result $x'$ in sublinear time and sample complexity and compared with the Fourier transform result $\hat{x}$, the following approximation guarantee holds:
 
\begin{align*}
    \| ( x' - \wh{x} )_S  \|_2^2 \leq \epsilon \| \wh{x}_{\bar{S}} \|_2^2 + \delta \| \wh{x} \|_1^2 
\end{align*}
with probability at least $9/10$. For a set $S \subseteq [n]$ and a vector $x \in \R^n$, we define  $x_S$ by setting if $i\in S$, $(x_S)_i = x_i$ and otherwise $(x_S)_i =0 $.

\begin{table}[!ht]\label{tab:1}
    \centering
    \begin{tabular}{|l|l|l|} \hline
        {\bf References} & {\bf Samples} & {\bf Time}  \\ \hline  
        \cite{hikp12a} & $\epsilon^{-1} k \log^2(n)  $ &  $\epsilon^{-1} k \log^2(n)$  \\ \hline 
        \cite{k17} & $\epsilon^{-1} k$ & $\epsilon^{-1} k \log^{2.1} (n) \log (R^*)$ \\ \hline
        Ours & $\epsilon^{-1} k \log (n)$ & $\epsilon^{-1} k \log (n)$ \\ \hline
    \end{tabular}
    \vspace{2mm}
\caption{Summary of the history of results
    }
\label{tab:history_of_results}
\end{table}

For this Fourier set query problem, there are two major prior works \cite{k17} and \cite{hikp12a}. The \cite{k17} studies the problem explicitly and \cite{hikp12a} implicitly provides a solution to Fourier set query, we will provide more details in the later paragraphs.
 
The work by \cite{k17} first explicitly define Fourier set query problem and studies it. \cite{k17} obtains an algorithm that has sample complexity $O(k/\epsilon)$ 
and running time $O(\epsilon^{-1} k \log^{2.1} (n) \log (R^*))$ for $\ell_2/\ell_2$ Fourier set query. Here, $R^*$ is an upper bound on the $\|\cdot\|_{\infty}$ norm of the vector.  In most applications, $R^*$ are considered $\poly(n)$. Our approach gives an algorithm with $O( \epsilon^{-1} k\log (n))$ running time. The  running time of our result has no dependence  on $\log R^*$, but our result do not achieve the optimal sample complexity.

The \cite{hikp12a} didn't study Fourier set query problem, instead they study Fourier sparse recovery problem. However, applying their algorithm \cite{hikp12a} to Fourier set query, it provides an algorithm with time complexity of $O( \epsilon^{-1} k \log^2 (n))$ and sample complexity of $O( \epsilon^{-1} k \log^2 (n)) $.

Our main contributions are summarized as follows:
\begin{itemize}
    \item We present a efficient algorithms for Fourier set query problem.
    \item We provide comprehensive theoretical guarantees to show the predominance of our algorithms over the existing algorithm.
\end{itemize}

{\bf Roadmap.}
We first present the related work about discrete fourier transform, continuous fourier transform and some applications of fourier transform in  Section~\ref{sec:relatedwork}. We define our problem and present our main theorem in Section~\ref{sec:fourier}. We present a high-level overview of our techniques in Section~\ref{sec:techniqueoverview}. We provide some definitions, notations and technique tools in Section \ref{sec:fourier_definitions_and_backgrounds}.  And as our main result in this paper, our algorithm (See Algorithm~\ref{alg:fourier_set_query_I}.) and the analysis about the correctness and complexity of it is given in Section~\ref{sec:mainresult}. Finally, we conclude our paper in  Section~\ref{sec:conclusion}.

\section{Related Work}\label{sec:relatedwork}

\paragraph{Discrete Fourier Transform}
For computational jobs, among the most crucial and often employed algorithms is the discrete Fourier transform (DFT).. There is a long line of works focus on sparse discrete Fourier
transforms. Results can be divided into two kinds: the first kind of results
choose sublinear measurements and achieve sublinear or linear recovery time. This kind of work  includes~\cite{gms05, hikp12a, hikp12b, iw13, ikp14, ik14, k16, k17, nsw19}.
 The second kind of results randomly choose measurements and prove that a generic recovery algorithm succeed with high probability. A common generic recovery algorithm that this kind of works used is $\ell_1$ minimization. These results prove the Restricted Isometry Property~\cite{crt06, rv08, b14}. Currently, the first kind of solutions have better theoretical guarantee in sample and time complexity. However, the second kind of algorithm has high success probabilities
and higher capability in practice.

\paragraph{Continuous Fourier Transform}
~\cite{sahgka13} studies sparse Fourier transforms on a continuous signals.
They apply a discrete sparse Fourier transform algorithm, followed by a hill-climbing method to optimize their solution into a reasonable range. 
 \cite{ps15} presents an algorithm whose sample complexity is only linear to $k$ and logarithmic in the signal-to-noise ratio. Their frequency resolution is suitable for robustly computing sparse continuous Fourier transforms. \cite{jls20} generalizes \cite{ps15} into high-dimensional setting. 
 \cite{ckps16} provide an algorithm that support the reconstruction of a signal without frequency gap. They present a solution to approximate the signal using a constant factor noise growth and takes samples polynomial in $k$ and logarithmic in the signal-to-noise ratio. Recently \cite{sswz22} improves the approximation ratio of \cite{ckps16}.

\paragraph{Application of Fourier Transform}
Fourier transformation has a wide application in many fields including physics, mathematics, signal processing, probability theory, statistics, acoustics, cryptography and so on.

Solving partial differential equations is one of the most important application of Fourier transformation.
Some differential equations are simpler to understand in the frequency domain because the action of differentiation in the time domain corresponds to the multiplication by the frequency. Additionally, frequency-domain multiplication is equivalent to convolution in the time domain \cite{mcg91}, \cite{pj01}, \cite{ffj98}.

Various applications of the Fourier transform include nuclear magnetic resonance (NMR) \cite{hdi97}, \cite{rabi38}, \cite{sks12} and other types of spectroscopy, such as infrared (FTIR) \cite{gp83}. In NMR, a free induction decay (FID) signal with an exponential shape is recorded in the time domain and Fourier-transformed into a Lorentzian line-shape in the frequency domain. Mass spectrometry and magnetic resonance imaging (MRI) both employ the Fourier transform. The Fourier transform is also used in quantum mechanics \cite{wm13}.

For the spectrum analysis of time-series \cite{sp10}, \cite{sl91}, the Fourier transform is employed. The Fourier transformation is often not applied to the signal itself in the context of statistical signal processing. It has been discovered in practice that it is best to simulate a signal by a function (or, alternatively, a stochastic process) that is stationary in the sense that its distinctive qualities are constant across all time, even though a genuine signal is in fact transitory. It has been discovered that taking the Fourier transform of the function's autocorrelation function is more advantageous for the analysis of signals since the Fourier transform of such a function does not exist in the conventional sense.


\section{Fourier set query}\label{sec:fourier}
 
In Section~\ref{sec:fsqproblem}, We define the problem we focus on. In Section \ref{sec:our_res}, we provide our main result.  


\subsection{Fourier set query problem}\label{sec:fsqproblem}
In this section, we give a formal definition of the main problem focused on.

\begin{definition}[Sample Complexity]\label{def:samplecomplexity}
Given a vector $x\in \C^n$, we say the sample complexity of an algorithm is $c$ (an Algorithm takes $c$ samples), when $c$ is the number of the coordinates used and $c\leq n$.
\end{definition}

\begin{definition}[Main problem]\label{def:fourier_set_query}
Given a vector $x \in \C^n$ and the $\hat{x}$ as the concrete Fourier transformation result, then for every $\epsilon, \delta \in (0,1)$ and $k\geq 1$, any $S \subseteq [n]$, $|S| = k$, the goal is to design an algorithm that 
\begin{itemize}
    \item takes samples from $x \in \C^n$ (note that we treat one entry of $x$ as one sample)
    \item takes some time to output a vector $x'\in \C^n$ such that
\begin{align*}
\| ( x' - \wh{x} )_S  \|_2^2 \leq \epsilon \| \wh{x}_{\bar{S}} \|_2^2 + \delta \| \wh{x} \|_1^2 
\end{align*}
\end{itemize}
We want to optimize both sample complexity (which is the number of coordinates we need to access in $x$), and also the running time.
\end{definition}

\subsection{Our Result}\label{sec:our_res}

We present our main theorem as follows:
\begin{theorem}[Main result]\label{thm:fourier_set_query_I_prelim}
Given a vector $x \in \C^n$ and the $\hat{x}$ as the concrete Fourier transformation result, then for every $\epsilon, \delta \in (0,1)$ and $k\geq 1$, any $S \subseteq [n]$, $|S| = k$, there exists an algorithm (Algorithm~\ref{alg:fourier_set_query_I}) that takes 
\begin{align*}
O(\epsilon^{-1} k \log (n/\delta)  )
\end{align*}
samples from $x$, runs in 
\begin{align*}
O(\epsilon^{-1} k \log (n/\delta) )
\end{align*}
time, and outputs a vector $x'\in \C^n$ such that
\begin{align*}
\| ( x' - \wh{x} )_S  \|_2^2 \leq \epsilon \| \wh{x}_{\bar{S}} \|_2^2 + \delta \| \wh{x} \|_1^2, 
\end{align*}
holds with probability at least $9/10$.
\end{theorem}

\section{Technique Overview}\label{sec:techniqueoverview}
In this section, we will give an overview about the technique methods used on the proof of our main result and the complexity analysis about time and sample (See Definition~\ref{def:samplecomplexity}.). At first, we will give an introduction about main functions and their time complexity as well as other properties used in our algorithm. And based on the functions, then we will give the analysis about the correctness of our algorithm where with probability at least $9/10$ it can finally produce a $x'$  which satisfies $$ \| ( x' - \wh{x} )_S  \|_2^2 \leq \epsilon \| \wh{x}_{\bar{S}} \|_2^2 + \delta \| \wh{x} \|_1^2 .$$

The analysis of total complexity comes last, with $O(\epsilon^{-1} k \log (n/\delta) )$ as the sample complexity (See Definition~\ref{def:samplecomplexity}) and $O(\epsilon^{-1} k \log (n/\delta) )$ as the time complexity. And then we can make sure the algorithm solve the problem (See Definition~\ref{def:fourier_set_query}.) with better performance compared to the prior works \cite{k17} and \cite{hikp12a} (See details in Table~\ref{tab:1}).

\paragraph{Technique I: \textsc{HashToBins}}
We use the same function \textsc{HashToBins} with the one in \cite{hikp12a}, which is one of the key part of the function EstimateValues. We can attain a $\hat{u}$, where the $\hat{u}_j$ for satisfies the following equation
\begin{align*}
\wh{u}_j = \sum_{h_{\sigma,b}(i) = j} \wh{(x-z)}_i (\wh{G'_{B,\delta,\alpha}})_{-o_{\sigma,b}(i)} \omega^{a \sigma i} \pm \delta \| \wh{x} \|_1.
\end{align*}
To help the analysis of the time complexity of our algorithm~\ref{alg:fourier_set_query_I}, by Lemma~\ref{lem:u_hashbin}, the time complexity of the function above is $O(\frac{B}{\alpha} \log (n/\delta) + \| \wh{z} \|_0 + \zeta \log (n/\delta) )$ with 
\begin{align*}
  \zeta = | \{ i \in \mathrm{supp}(\wh{z}) ~|~ E_{\mathrm{off}}(i) \} |.  
\end{align*}

\paragraph{Technique II: EstimateValues}
EstimateValues is an key function in loop (See Section~\ref{sec:fourier_iterative_loop_analysis}). By using this function, we attain the new set $T_i$ and the new value $\hat{w}^{(i)}$ to update $S_{i}$ by
\begin{align*}
  S_{i+1} \leftarrow S_i \backslash T_i  ,
\end{align*}
and $\wh{z}^{(i+1)}$ by 
\begin{align*}
  \wh{z}^{(i+1)} \leftarrow \wh{z}^{(i)} + \wh{w}^{(i)}.  
\end{align*}

\paragraph{Technique III: Query Set $S$}
We use $S$ as the query set and $S_i$ is the set attained by updating $S$ with $i-1$ iterations. And we use $k_i=k\gamma^{i-1}$ where $\gamma\leq\frac{1}{1000}$ and $k\geq 1$.

We demonstrate that we can compress $S_i$ to a small enough value where $|S_{i}| \leq k_{i}$. Due to reason that $S_i$ is a query set, the above sentence can be said as that we can finish the query of all the elements in $S$ with a large enough number of the iterations.  

In the proof of above statement, we bring some properties about $t$ as follows (See Details in Definition~\ref{def:cll}):
\begin{enumerate}
    \item ``Collision'' 
    \item ``Large offset'' 
    \item ``Large noise'' 
\end{enumerate}
Given a vector $x$ and  $t\in [n]$ as a coordinate of it, we also give the definition about ``well-isolated'' based on concepts above. And then we can prove that with probability at least $1-a_i$, we can have  $t$ is ``well-isolated''.

Based on the statement above, we can have small enough $|S_i|$ by  $|S_{i}|\leq k_{i}$ and a large enough $R$ here.

\paragraph{Technique IV: Correctness and Complexity}
By the upper bound of $\| \wh{x}_{\ov{S}_{i+1}}^{(i+1)} \|_2^2$ which we attain in Section~\ref{sec:fourier_iterative_loop_analysis}. We can demonstrate the error can satisfy the requirement in the problem. With probability $10 a_i/\gamma$, we can have
\begin{align*}
    \| \wh{x}_{\ov{S}_{i+1}}^{(i+1)} \|_2^2 \leq (1+\epsilon_i)\| \wh{x}^{(i)}_{\ov{S}_i} \|_2^2 + \epsilon_i \delta^2 n \| \wh{x} \|_1^2.
\end{align*}

Then we can demonstrate
\begin{align*}
 \| \wh{x}_S - \wh{z}^{(R+1)} \|_2^2\leq \epsilon (\| \wh{x}_{\ov{S}} \|_2^2 + \delta^2 n \| \wh{x} \|_1^2).
\end{align*}

Notice that the $\wh{z}^{R+1}$ is the output of our Algorithm~\ref{alg:fourier_set_query_I} which is also the $x'$ in our problem (See Definition~\ref{def:fourier_set_query}). The above inequalities demonstrate that the Algorithm~\ref{alg:fourier_set_query_I} constructed by us can output a $x'$ which satisfies 
\begin{align*}
    \| (\wh{x}-x')_S  \|_2^2 \leq \epsilon \| \wh{x}_{\bar{S}} \|_2^2 + \delta \| \wh{x} \|_1^2
\end{align*}
with succeed probability 9/10.
And we attain sample complexity and time complexity by 
\begin{align*}
   \sum_{i=1}^R (B_i / \alpha_i) \log (n/\delta) = \epsilon^{-1} k \log (n/\delta).
\end{align*} 


\section{Preliminary}
\label{sec:fourier_definitions_and_backgrounds} 

In this section, we first present some definitions and background for Fourier transform in Section \ref{sec:notation}. We introduce some technical tools in Section \ref{sec:tec_tool}. 
Then we introduce spectrum permutations and filter functions in Section~\ref{sec:fourier_permutation_and_filter_function}. They are used as hashing schemes in the Fourier transform literature.
In Section~\ref{sec:fourier_collison_offset_noise}, we introduce collision events. large offset events, and large noise events.

\subsection{Notations}\label{sec:notation}
We use $\i$ to denote $\sqrt{-1}$. Note that $e^{\i \theta} = \cos(\theta) + \i \sin(\theta)$. For any complex number $z\in \C$, we have $z = a + \i b$, where $a,b \in \R$. We define the complement of $z$ as $\ov{z} = a - \i b$. We define $|z| = \sqrt{ z\ov{z} } = \sqrt{a^2 +b^2}$. For any complex vector $x\in \C^n$, we use $\supp(x)$ to denote the support of $x$, and then $\| x \|_0 = |\supp(x)|$. We define $\omega = e^{2\pi \i /n}$, which is the $n$-th unitary root i.e. $\omega^n=1$.

The discrete convolution of functions $f$ and $g$ is given by,
\begin{align*}
(f * g) [n] = \sum_{m = -\infty}^{+\infty} f[m] g[n-m]
\end{align*}
For a complex vector $x \in \C^n$, we use $\wh{x} \in \C^n$ to denote its Fourier spectrum,
\begin{align*}
\wh{x}_i = \frac{1}{\sqrt{n}} \sum_{j=1}^n e^{-2\pi \i i j/ n} x_j, \forall i\in [n].
\end{align*}
Then the inverse transform is
\begin{align*}
x_j = \frac{1}{\sqrt{n}} \sum_{i=1}^n e^{2\pi \i i j/ n} \wh{x}_i, \forall j \in [n].
\end{align*}
We define
\begin{align*}
\Err(x,k) := \min_{k\text{-sparse}~y} \| x - y \|_2.
\end{align*}
We define $x_S$ as a vector by setting if $i \in S$, $(x_S)_i = x_i$  and otherwise $(x_S)_i =0 $, for  a vector $x \in \R^n$ and a set $S \subseteq [n]$.

\subsection{Technical Tools}\label{sec:tec_tool}

We show several technical tools and some lemmas in prior works we used in the following section.
\begin{lemma}[Markov's inequality]\label{lem:markov'sinequality}
If $X$ is a nonnegative random variable and $a > 0$, then the probability that $X$ is at least a is at most the expectation of $X$ divided by $a$:
\begin{align*}
\Pr[X\geq a]\leq \frac{\E(X)}{a}.
\end{align*}

Let $a=\wt{a}\cdot \E(X)$ (where $\wt{a}>0$); then we can rewrite the previous inequality as
\begin{align*}
\Pr[X\geq \wt{a}\cdot \E(X)]\leq \frac{1}{\wt{a}} 
\end{align*}
\end{lemma}

The following two lemmas of complex number are standard. We prove the following two lemmas for the completeness of the paper.

\begin{lemma}\label{lem:expectation_pairwise}
Given a fixed vector  $x \in \R^n$ and a pairwise independent random variable $\sigma_i$ where $\sigma_i = \pm 1$ with probability $1/2$ respectively. Then we have:

\begin{align*}
    \E_{\sigma} [(\sum_{i=1}^{n} \sigma_i x_i)^2] = \| x \|_2^2
\end{align*}
\end{lemma}
\begin{proof}
We have:
\begin{align*}
    &~\E_{\sigma} [(\sum_{i=1}^{n} \sigma_i x_i)^2] \\
    = &~\E[\sum_{i=1}^{n} \sigma_i^2 x_i^2] + \E[\sum_{i \neq j} \sigma_i x_i \sigma_j x_j] \\
    =&~ \E[\sum_{i=1}^{n} \sigma_i^2 x_i^2] + \sum_{i \neq j} \E[\sigma_i \sigma_j] x_i x_j \\
    =&~ \E[\sum_{i=1}^{n} \sigma_i^2 x_i^2] + \sum_{i \neq j} \E[\sigma_i] \cdot \E[ \sigma_j ] x_i x_j \\
    = &~\E[\sum_{i=1}^{n} \sigma_i^2 x_i^2] + 0 \\
    =&~ \| x \|_2^2
\end{align*}
where the first step comes from the linearity of expectation, the second step follows the linearity of expectation, the third step $\sigma_i$ is a pairwise independent random variable, the fourth step follows that $\E[\sigma_i]=0$ , and the final step comes from the definition of $\|\cdot\|_2$ and $\sigma_i^2 = 1$.
\end{proof}

\begin{lemma}\label{lem:complex_vector_expectation}
Let $a \sim [n]$ uniformly at random. Given a fixed vector  $x \in \C^n$ and $\omega^{\sigma a i}$, then we have:
\begin{align*}
    \E_{a} [|\sum_{i=1}^{n} x_i \omega^{\sigma a i} |^2] = \| x \|_2^2
\end{align*}
\end{lemma}

\begin{proof}
For any fixed $i\in [n]$, we have the inequality as follows
\begin{align}\label{eq:fact3.6}
\E_{a}[\omega^{ a i}] 
 = &~ \frac{1}{n} \sum_{a=1}^{n} \omega^{ a i} 
    = \frac{1}{n} \cdot \frac{1 - \omega^{ni}}{1 - \omega^{i}} 
    =~ 0
\end{align}
where the first step comes from geometric sum, and the second step comes from
We have:
\begin{align*}
    &~\E_{a} [|\sum_{i=1}^{n} x_i \omega^{\sigma a i}|^2]  \\
    = &~ \E_{a}[(\sum_{i=1}^{n} x_i \omega^{\sigma a i}) (\sum_{i=1}^{n} \bar{x}_i \omega^{-\sigma a i})   ] \\
    = &~\E_{a}[\sum_{i=1}^{n}  x_i \bar{x}_i ] + \E_{a}[\sum_{i \neq j}  x_i \omega^{\sigma a i}  \bar{x}_j \omega^{-\sigma a j}] \\
    = &~\E_{a}[\sum_{i=1}^{n}  x_i \bar{x}_i ] + \sum_{i \neq j} \E_{a}[  \omega^{\sigma a (i-j)}] x_i \bar{x}_j \\
    = &~\E_{a}[\sum_{i=1}^{n}  x_i \bar{x}_i ] + 0 \\
    =&~ \| x \|_2^2
\end{align*}
where the first step follows that for a complex number $z$, $|z|^2 = z \bar{z}$,
the second step follows the linearity of expectation,
the third step follows the linearity of expectation,
where the fourth step follows Eq.\ref{eq:fact3.6},
and the final step comes from the definition of $\|\cdot\|_2$.
\end{proof}

\subsection{Permutation and filter function}\label{sec:fourier_permutation_and_filter_function}
We use the same (pseudorandom) spectrum permutation as \cite{hikp12a},
\begin{definition}
Suppose $\sigma^{-1}$ exists mod $n$. We define the permutation $P_{\sigma,a,b}$ by
\begin{align*}
(P_{\sigma,a,b} x)_i = x_{\sigma(i-a)} e^{-2\pi \i \sigma b i /n }.
\end{align*}
\end{definition}
We also define $\pi_{\sigma,b} = \sigma(i-b) \pmod n$. Then we have
\begin{claim}
We have that
$$\wh{P_{\sigma,a,b} x}_{\pi_{\sigma,b}(i)} = \wh{x}_i e^{-2\pi \i \sigma a i /n}.$$
\end{claim}

$h_{\sigma,b}(i)$ is defined as the ``bin'' with the mapping of frequency $i$ onto. We define $o_{\sigma,b}(i)$ as the ``offset''. We formally define them as follows:
 
\begin{definition}
Let the hash function be defined as 
$$h_{\sigma,b}(i): = \mathrm{round}(\frac{\pi_{\sigma,b}(i) B }{n}).$$
\end{definition}

\begin{definition}
Let the offset function be defined as
$$o_{\sigma,b}(i): = \pi_{\sigma,b}(i) - h_{\sigma,b}(i)\frac{ n}{B}.$$
\end{definition}

We use the same filter function as \cite{hikp12a,ps15,ckps16},
\begin{definition}
Given parameters $B\geq 1$, $\delta >0$, $\alpha >0$. We say that $(G,\wh{G}') = (G_{B,\delta,\alpha}, \wh{G}'_{B,\delta,\alpha}) \in \R^n$ is a filter function 
if it satisfies the following properties: 
\begin{enumerate}
    \item $|\supp(G)| = O( \alpha^{-1} B \log (n/\delta))$.
    \item if $|i| \leq (1-\alpha) n / (2B)$, $\wh{G}'_i = 1$.
    \item if $|i| \geq n / (2B)$, $\wh{G}'_i = 0$.
    \item for all $i$, $\wh{G}'_i \in [0,1]$.
    \item $\left\| \wh{G}' - \wh{G} \right\|_{\infty} < \infty$.
\end{enumerate}
\end{definition}

\subsection{Collision event, large offset event, and large noise event}\label{sec:fourier_collison_offset_noise}

We use three types of events defined in \cite{hikp12a} as basic building blocks for analyzing Fourier set query algorithms. 
For any $i\in S$, we define three types of events associated with $i$ and $S$ and defined over the probability space induced by $\sigma$ and $b$:
\begin{definition}[Collision, large offset, large noise]\label{def:cll}
The definition of three events are given as follow:
\begin{itemize}
\item We say ``Large offset'' event $E_{\mathrm{off}}(i)$ holds if $$n (1-\alpha)/ (2B) \leq |o_{\sigma,b}(i)|.$$
\item We say ``Large noise'' event $E_{\mathrm{noise}}(i)$  holds if $$ (\alpha B)^{-1}\cdot\Err^2( \wh{x}', k ) \leq \E\left[ \left\| \wh{x}'_{ h^{-1}_{\sigma,b} ( h_{\sigma,b}(i) ) \backslash S } \right\|_2^2 \right].$$
\item We say ``Collision'' event $E_{\mathrm{coll}}(i)$ holds if $$h_{\sigma,b}(i) \in h_{\sigma,b}(S \backslash \{i\} ).$$
\end{itemize}

\begin{definition}[Well-isolated]\label{def:well-isolated}
For a vector $x\in \R^n$, we say a coordinate $t\in [n]$ is ``well isolated'' when none of  ``Collision'' event,  ``Large offset'' and ``Large noise'' event holds.
\end{definition}
\end{definition}

\begin{claim}[Claim 3.1 in \cite{hikp12a}]\label{cla:claim_3.1_hikp12a}
For all $i\in S$, we have
$$\Pr[E_{\mathrm{coll}}(i) ] \leq 4 \frac{|S| }{B}.$$
\end{claim}

\begin{claim}[Claim 3.2 in \cite{hikp12a}]\label{cla:claim_3.2_hikp12a}
For all $i\in S$, we have
$$\Pr[E_{\mathrm{off}}(i) ] \leq \alpha.$$
\end{claim}

\begin{claim}[Claim 4.1 in \cite{hikp12a}]\label{cla:claim_4.1_hikp12a}
For any $i\in S$, the event $E_{\mathrm{noise} (i) }$ holds with probability at most $4 \alpha$
$$\Pr[ E_{\mathrm{noise} (i) } ] \leq 4 \alpha.$$
\end{claim}

\begin{lemma}[Lemma 4.2 in \cite{hikp12a}]\label{lem:lemma_4.2_hikp12a}
With $B$ divide $n$, $a$ uniformly sampled from $[n]$ and the others without limitation in 
\begin{align*}
\wh{u} = \textsc{HashToBins}(P_{\sigma,a,b}, \alpha, \wh{z},B, \delta, x).
\end{align*}
With all of $E_{\mathrm{off}}(i)$, $E_{\mathrm{coll}}(i)$ and $E_{\mathrm{noise}}(i)$ not holding and $j = h_{\sigma,b}(i)$, we have for all $i\in[n]$, 
\begin{align*}
\E \left[ \left| \wh{x}'_i e^{- \frac{2 \pi \i}{n} a \sigma i} \right|^2 -\wh{u}_j \right] \leq 2 \frac{\rho^2}{\alpha B}.
\end{align*}
\end{lemma}

\begin{lemma}[Lemma 3.3 in \cite{hikp12a}]\label{lem:u_hashbin}
Suppose $B$ divides $n$. The output $\wh{u}$ of $\textsc{HashToBins}$ satisfies
\begin{align*}
\wh{u}_j = \sum_{h_{\sigma,b}(i) = j} \wh{(x-z)}_i (\wh{G'_{B,\delta,\alpha}})_{-o_{\sigma,b}(i)} \omega^{a \sigma i} \pm \delta \| \wh{x} \|_1.
\end{align*}
 
Let $$\zeta := | \{ i \in \mathrm{supp}(\wh{z}) ~|~ E_{\mathrm{off}}(i) \} |.$$ The running time of \textsc{HashToBins} is $$O(\frac{B}{\alpha} \log (n/\delta) + \| \wh{z} \|_0 + \zeta \log (n/\delta) ).$$
\end{lemma}

\section{Analysis on  Fourier Set Query Algorithm}\label{sec:mainresult}
In this section, we will give an total analysis about our Algorithm~\ref{alg:fourier_set_query_I}. First, we will provide the iterative loop analysis which is the main part of our main function \textsc{FourierSetQuery} in Section~\ref{sec:fourier_iterative_loop_analysis}. By this analysis, we demonstrate an important property of the Algorithm~\ref{alg:fourier_set_query_I} in Section \ref{sec:induction}. 
In Section~\ref{sec:fourier_main_result}, we prove the the correctness of the algorithm. We also provide the analysis of the complexity (sample and time) of Algorithm~\ref{alg:fourier_set_query_I}. Then we can give an satisfying answer to the problem (See Definition~\ref{def:fourier_set_query}) with Algorithm~\ref{alg:fourier_set_query_I} attained by us whose performance (on sample and time complexity) is better than prior works (See Table~\ref{tab:1}).

\subsection{Iterative loop analysis}\label{sec:fourier_iterative_loop_analysis}
Iterative loop analysis for Fourier set query is more tricky than the classic set query, because in the Fourier case, hashing is not perfect, in the sense that by using spectrum permutation and filter function (as the counterpart of hashing techniques), one coordinate can non-trivially contribute to multiple bins. We give iterative loop induction in Lemma~\ref{lem:iterative_loop_analysis_fourier_set_query_I}.

\begin{lemma} \label{lem:wellisolated}
Given a vector $x\in \R^n$, $\gamma\leq 1/1000$, $\alpha_i=1/(200 i^3)$, for a coordinate $t\in [n]$ and each $i\in [R]$, with probability at least 
$1-6\alpha_i$,
We say that $t$ is ``well isolated'' (See Definition~\ref{def:well-isolated}).
\end{lemma}
\begin{proof}
{\bf Collision.}
Using Claim~\ref{cla:claim_3.1_hikp12a}, for any $t \in S_{i}$, the event $E_{\mathrm{coll}}(t)$ holds with probability at most 
\begin{align*}
4 |S_{i} | / B_i 
\leq&~ \frac{4 k_i}{ C k_i / (\alpha_i^2 \epsilon_i)  } \\
=&~ 4 \alpha_i^2 \epsilon_i /C \\
\leq&~ \alpha_i,
\end{align*}
where the first step follows from the definition of $B_i$ and the assumption on $ |S_i|$, the second step is straightforward, the third step follows from the definition of $\eps_i$, $\alpha_i$, and $C$. 

It means
\begin{align*}
\Pr_{\sigma,b} \left[ E_{\mathrm{coll}}(t) \right] \leq \alpha_i.
\end{align*}

{\bf Large offset.}
Using Claim~\ref{cla:claim_3.2_hikp12a}, for any $t \in S_i$, the event $E_{\mathrm{off}}(t)$ holds with probability at most $\alpha_i$, i.e.
\begin{align*}
\Pr_{\sigma,b} \left[ E_{\mathrm{off}}(t) \right] \leq \alpha_i.
\end{align*}

{\bf Large noise.}
Using Claim~\ref{cla:claim_4.1_hikp12a}, for any $t \in S_i$,
\begin{align*}
\Pr_{\sigma,b}[ E_{\mathrm{noise}}(t) ] \leq 4 \alpha_i.
\end{align*}

By a union bound over the above three events, we have $t$ is ``well isolated'' with probability at least $1-6\alpha_i$.
\end{proof}
\begin{lemma}\label{lem:fourier_set_query_I_noise_00}
Given parameters $C \geq 1000$, $\gamma \leq 1/1000$. For any $k\geq 1 , \epsilon \in (0,1)$, $R\geq 1$.
For each $i \in [R]$, we define
\begin{align*}
k_i :=&~ k \gamma^{i-1}, \\
\epsilon_i :=&~  \epsilon (10\gamma)^i,\\
\alpha_i :=&~ 1/(200 i^3) ,\\
B_i :=&~ C \cdot k_i/ (\alpha_i^2 \epsilon_i ).
\end{align*}
For each $i\in [R]$: If for all $j \leq [i-1]$ we have
\begin{enumerate}
    \item $\supp( \wh{w}^{(j)} ) \subseteq S_j$.
    \item $| S_{j+1} | \leq k_{j+1}$.
    \item $\wh{z}^{(j+1)} = \wh{z}^{(j)} + \wh{w}^{(j)}$.
    \item $\wh{x}^{(j+1)} = \wh{x} - \wh{z}^{(j+1)}$.
    \item $\| \wh{x}_{\ov{S}_{j+1}}^{(j+1)} \|_2^2 \leq (1+\epsilon_j) \| \wh{x}_{\ov{S}_{j}}^{(j)} \|_2^2 + \epsilon_j \delta^2 n \| \wh{x} \|_1^2$.
\end{enumerate}

Then, with probability $1- 10 \alpha_i / \gamma$, we have
\begin{align*}
 |S_{i+1}|\leq k_{i+1}. 
\end{align*}

\end{lemma}

\begin{proof}

We consider a particular step $i$. We can condition on $ | S_i | \leq k_i$.

By Lemma~\ref{lem:wellisolated},  we have $t$ is ``well isolated'' with probability at least $1-6\alpha_i$.

Therefore, each $t\in S_i$ lies in $T_i$ with probability at least $1- 6 \alpha_i$.  We have Then by Markov's inequality (See Lemma~\ref{lem:markov'sinequality}) and assumption in the statement, we have 
\begin{align}
    |S_i \backslash T_i| \leq \gamma k_{i} \label{eq:s_sub_t_leq_gamma}
\end{align}
with probability $1- 6 \alpha_i/\gamma$. Then we know that
\begin{align*}
| S_{i+1} | 
=&~ | S_i \backslash T_i| \\
\leq&~ \gamma k_{i} \\
\leq&~ k_{i+1}.
\end{align*}
where the first step follows from the definition of $S_{i+1} = S_i \backslash T_i$, the second step follows from Eq.~\eqref{eq:s_sub_t_leq_gamma}, the third step follows from the definition of $ k_{i}$ and $ k_{i+1}$.

\end{proof}

\begin{lemma}\label{lem:fourier_set_query_I_noise_0}
Given parameters $C \geq 1000$, $\gamma \leq 1/1000$. For any $k\geq 1 , \epsilon \in (0,1)$, $R\geq 1$.
For each $i \in [R]$, we define
\begin{align*}
k_i :=&~ k \gamma^{i-1}, \\
\epsilon_i :=&~  \epsilon (10\gamma)^i,\\
\alpha_i :=&~ 1/(200 i^3) ,\\
B_i :=&~ C \cdot k_i/ (\alpha_i^2 \epsilon_i ).
\end{align*}
For each $i\in [R]$: If for all $j \leq [i-1]$ we have
\begin{enumerate}
    \item $\supp( \wh{w}^{(j)} ) \subseteq S_j$.
    \item $| S_{j+1} | \leq k_{j+1}$.
    \item $\wh{z}^{(j+1)} = \wh{z}^{(j)} + \wh{w}^{(j)}$.
    \item $\wh{x}^{(j+1)} = \wh{x} - \wh{z}^{(j+1)}$.
    \item $\| \wh{x}_{\ov{S}_{j+1}}^{(j+1)} \|_2^2 \leq (1+\epsilon_j) \| \wh{x}_{\ov{S}_{j}}^{(j)} \|_2^2 + \epsilon_j \delta^2 n \| \wh{x} \|_1^2$.
\end{enumerate}

Then, with probability $1- 10 \alpha_i / \gamma$, we have
\begin{align*}
 \Pr \left[ \left\| \wh{x}^{(i)}_{T_i} - \wh{w}^{(i)} \right\|_2^2 \leq  \frac{\epsilon_i}{20} ( \| \wh{x}^{(i)}_{\ov{S}_i} \|_2^2 + \delta^2 n \| \wh{x} \|_1^2 ) \right] \geq 1- \alpha_i.
\end{align*}

\end{lemma}

\begin{proof}

We define $\rho^{(i)}$ and $\mu^{(i)}$ as follows
\begin{align}\label{eq:mu_def}
\rho^{(i)} = & ~ \left\| \wh{x}^{(i)}_{\ov{S}_i} \right\|_2^2 + \delta^2 n \| \wh{x} \|_1^2 , \notag \\
\mu^{(i)}  = & ~ \frac{\epsilon_i}{k_i} \left( \left\| \wh{x}^{(i)}_{\ov{S}_i} \right\|_2^2 + \delta^2 n \| \wh{x} \|_1^2 \right).
\end{align}

For a fixed $t\in S_i$, let $j = h_{\sigma,b}(t)$.
By Lemma~\ref{lem:u_hashbin}, we have
\begin{align}\label{eq:u_hashbin}
\wh{u}_j - \wh{x}_{t}^{(i)} \omega^{a\sigma t} = \sum_{t'\in T_i} \wh{G}'_{-o_{\sigma}(t')} \wh{x}_{t'}^{(i)} \omega^{a\sigma t'} \pm \delta \| \wh{x} \|_1
\end{align}

For each $t \in S_i$, we define set $Q_{i,t} = h^{-1}_{\sigma,b}(j) \backslash \{t\}$.
Let $T_i$ be the set of coordinates $t \in S_i$ such that $Q_{i,t} \cap S_i = \emptyset$. Then it is easy to observe that
\begin{align*}
& ~ \sum_{t \in T_i} \left| \sum_{t' \in Q_{i,t} } \wh{G}'_{-o_{\sigma}(t')} \wh{x}_{t'}^{(i)} \omega^{a\sigma t'} \right|^2 \\
= & ~ \sum_{t \in T_i} \left| \sum_{t' \in Q_{i,t} \backslash S_i} \wh{G}'_{-o_{\sigma}(t')} \wh{x}_{t'}^{(i)} \omega^{a\sigma t'} \right|^2 \\
\leq & ~ \sum_{t \in S_i} \left| \sum_{t' \in Q_{i,t} \backslash S_i} \wh{G}'_{-o_{\sigma}(t')} \wh{x}_{t'}^{(i)} \omega^{a \sigma t'} \right|^2
\end{align*}
 where the first step comes from $Q_{i,t} \cap S_i = \emptyset$, and the second step follows that $T_i \subseteq S_i$.

We can calculate the expectation of $\| \wh{x}^{(i)}_{T_i} - \wh{w}^{(i)} \|_2^2$.

We first demonstrate that $$ \E_{\sigma,a,b} \left[ \left\| \wh{x}^{(i)}_{T_i} - \wh{w}^{(i)} \right\|_2^2 \right]=\E_{\sigma,a,b}\left[ \sum_{t\in T_i} | \wh{x}_{t}^{(i)} - \wh{u}_{h_{\sigma,b}(t)} \omega^{-a \sigma  t} |^2 \right].$$
then get the upper bound of $$\E_{\sigma,a,b}\left[ \sum_{t\in T_i} | \wh{x}_{t}^{(i)} - \wh{u}_{h_{\sigma,b}(t)} \omega^{-a \sigma  t} |^2 \right]$$.

We have
\begin{align*}
 \E_{\sigma,a,b} \left[ \left\| \wh{x}^{(i)}_{T_i} - \wh{w}^{(i)} \right\|_2^2 \right] 
= & ~ \E_{\sigma,a,b}\left[ \sum_{t\in T_i} | \wh{x}_{t}^{(i)} - \wh{w}_t^{(i)} |^2 \right] \\
= &~  \E_{\sigma,a,b}\left[ \sum_{t\in T_i} | \wh{x}_{t}^{(i)} - \wh{u}_{h_{\sigma,b}(t)} \omega^{-a \sigma  t} |^2 \right] \\
= &~  \E_{\sigma,a,b}\left[ \sum_{t\in T_i} |  \wh{x}_{t}^{(i)}  \omega^{a \sigma  t} -\wh{u}_{h_{\sigma,b}(t)} |^2 \right] 
\end{align*}
 where the first step follows that summation over $T_i$, the second step comes from the definition of $\wh{w}_t^{(i)}$(in Line~\ref{alg:wh_w_t_def} in Algorithm~\ref{alg:fourier_set_query_I}), the third step follows that $$| \wh{x}_{t}^{(i)} - \wh{u}_{h_{\sigma,b}(t)} \omega^{ - a \sigma  t} | = |\omega^{- a \sigma  t}| \cdot | \wh{x}_{t}^{(i)}  \omega^{a \sigma  t} - \wh{u}_{h_{\sigma,b}(t)}   |$$ and $|\omega^{- a \sigma  t}|  = 1$,
the fourth step comes from Eq.~\eqref{eq:u_hashbin}.

And then we have
\begin{align*}
& ~ \E_{\sigma,a,b} \left[ \left\| \wh{x}^{(i)}_{T_i} - \wh{w}^{(i)} \right\|_2^2 \right] \\
= &~  \E_{\sigma,a,b}\left[ \sum_{t\in T_i} |  \wh{x}_{t}^{(i)}  \omega^{a \sigma  t} -\wh{u}_{h_{\sigma,b}(t)} |^2 \right] \\
\leq &  ~ \sum_{t \in S_i} 2 \E_{\sigma,a,b} \left[ \left| \sum_{t' \in Q_{i,t} \backslash S_i} \wh{G}'_{-o_{\sigma}(t')}{\wh{x}_{t'}^{(i)}} \omega^{a\sigma t'} \right|^2  \right] + \delta^2 \| \wh{x} \|_1^2  \\
\leq &  ~ \sum_{t \in S_i} 2 \E_{\sigma,b} \left[ \sum_{t' \in Q_{i,t} \backslash S_i} \left|  \wh{G}'_{-o_{\sigma}(t')}{\wh{x}_{t'}^{(i)}} \right|^2  \right] + \delta^2 \| \wh{x} \|_1^2  \\
= &  ~ \sum_{t \in S_i} 2 \E_{\sigma,b} \left[ \sum_{t' \in \bar{S}_i} \textbf{1}(t' \in Q_{i,t} \backslash S_i) \cdot \left|  \wh{G}'_{-o_{\sigma}(t')}{\wh{x}_{t'}^{(i)}} \right|^2  \right] + \delta^2 \| \wh{x} \|_1^2  \\
\leq & ~ \sum_{t\in S_i} (\frac{1}{B_i} \| \wh{x}^{(i)}_{\ov{S}_i} \|_2^2 + \delta^2 \| \wh{x} \|_1^2)\\
\leq & ~ \frac{|S_i|}{B_i} \| \wh{x}^{(i)}_{\ov{S}_i} \|_2^2 + \delta^2 |S_i| \cdot \| \wh{x} \|_1^2 \\
\leq & ~ \frac{\epsilon_i \alpha_i^2}{C} \| \wh{x}^{(i)}_{\ov{S}_i} \|_2^2 + \delta^2 |S_i| \cdot \| \wh{x} \|_1^2,
\end{align*}

 where  the first step follows the equation above, the second step follows Lemma~\ref{lem:complex_vector_expectation}, 
 the third step follows from expanding the squared sum, the fourth step follows that if $A_1 \subseteq A_2$, we have $$\sum_{i \in A_1}f(i) = \sum_{i \in A_2}\textbf{1}(i \in A_1) f(i),$$
the fifth step follows for two pairwise independent random variable $t$ and $t'$, we have $h_{\sigma,b}(t) = h_{\sigma,b}(t')$ holds with probability at most $1/B_i$, the sixth step comes from the summation over $S_i$, and the last step follows from $|S_i| \leq k_i$ and $B_i = C \cdot k_i / (\alpha_i^2 \epsilon_i)$. 

Then, using Markov's inequality, we have,
\begin{align*}
 \Pr \left[ \left\| \wh{x}^{(i)}_{T_i} - \wh{w}^{(i)} \right\|_2^2 \geq \frac{\epsilon_i \alpha_i}{C} \| \wh{x}^{(i)}_{\ov{S}_i} \|_2^2 + \delta^2 \frac{|S_i|}{\alpha_i} \| \wh{x} \|_1^2 \right] \leq \alpha_i.
\end{align*}

Note that
\begin{align*}
\frac{\epsilon_i \alpha_i}{C} \| \wh{x}^{(i)}_{\ov{S}_i} \|_2^2 + \delta^2 \frac{|S_i|}{\alpha_i} \| \wh{x} \|_1^2 
\leq & ~ \frac{\epsilon_i}{C} \| \wh{x}^{(i)}_{\ov{S}_i} \|_2^2 + \delta^2 \frac{|S_i|}{\alpha_i} \| \wh{x} \|_1^2 \\
\leq & ~ \frac{\epsilon_i}{C} \| \wh{x}^{(i)}_{\ov{S}_i} \|_2^2 + \frac{\epsilon_i}{C} \delta^2 B_i \| \wh{x} \|_1^2 \\
\leq & ~ \frac{\epsilon_i}{C} \| \wh{x}^{(i)}_{\ov{S}_i} \|_2^2 + \frac{\epsilon_i}{C} \delta^2 n \| \wh{x} \|_1^2 \\
\leq &~ \frac{\epsilon_i}{20} ( \| \wh{x}^{(i)}_{\ov{S}_i} \|_2^2 + \delta^2 n \| \wh{x} \|_1^2 ),
\end{align*}

where the first step follows by $\alpha_i \leq 1$, the second step follows by $|S_i| \leq k_i = \epsilon_i B_i \alpha_i^2 / C$, the third step follows by $B_i \leq n$, the last step follows by $C \geq 1000$.

Thus, we have
\begin{align*}
 \Pr \left[ \left\| \wh{x}^{(i)}_{T_i} - \wh{w}^{(i)} \right\|_2^2 \leq  \frac{\epsilon_i}{20} ( \| \wh{x}^{(i)}_{\ov{S}_i} \|_2^2 + \delta^2 n \| \wh{x} \|_1^2 ) \right] \geq 1- \alpha_i.
\end{align*}

\end{proof}

\begin{lemma}\label{lem:iterative_loop_analysis_fourier_set_query_I}
Given parameters $C \geq 1000$, $\gamma \leq 1/1000$. For any $k\geq 1 , \epsilon \in (0,1)$, $R\geq 1$.
For each $i \in [R]$, we define
\begin{align*}
k_i :=&~ k \gamma^{i-1}, \\
\epsilon_i :=&~  \epsilon (10\gamma)^i,\\
\alpha_i :=&~ 1/(200 i^3) ,\\
B_i :=&~ C \cdot k_i/ (\alpha_i^2 \epsilon_i ).
\end{align*}
For each $i\in [R]$: If for all $j \leq [i-1]$ we have
\begin{enumerate}
    \item $\supp( \wh{w}^{(j)} ) \subseteq S_j$.
    \item $| S_{j+1} | \leq k_{j+1}$.
    \item $\wh{z}^{(j+1)} = \wh{z}^{(j)} + \wh{w}^{(j)}$.
    \item $\wh{x}^{(j+1)} = \wh{x} - \wh{z}^{(j+1)}$.
    \item $\| \wh{x}_{\ov{S}_{j+1}}^{(j+1)} \|_2^2 \leq (1+\epsilon_j) \| \wh{x}_{\ov{S}_{j}}^{(j)} \|_2^2 + \epsilon_j \delta^2 n \| \wh{x} \|_1^2$.
\end{enumerate}
Then, with probability $1- 10 \alpha_i / \gamma$, we have
\begin{enumerate}
    \item $\supp( \wh{w}^{(i)} ) \subseteq S_i$.
    \item $| S_{i+1} | \leq k_{i+1}$.
    \item $\wh{z}^{(i+1)} = \wh{z}^{(i)} + \wh{w}^{(i)}$.
    \item $\wh{x}^{(i+1)} = \wh{x} - \wh{z}^{(i+1)}$.
    \item $\| \wh{x}_{\ov{S}_{i+1}}^{(i+1)} \|_2^2  \leq (1+\epsilon_i) \| \wh{x}_{\ov{S}_{i}}^{(i)} \|_2^2  + \epsilon_i \delta^2 n \| \wh{x} \|_1^2$.
\end{enumerate}
\end{lemma}

\begin{proof}
We will prove the five results one by one.

{\bf Part 1.}

Follows from Line \ref{ln:supp_w_S} in the Algorithm \ref{alg:fourier_set_query_I}, we have that
\begin{align*}
    \supp( \wh{w}^{(i)} ) \subseteq S_i.
\end{align*}

{\bf Part 2.}

By Lemma \ref{lem:fourier_set_query_I_noise_00}, we have that
\begin{align*}
| S_{i+1} | \leq k_{i}. 
\end{align*}

{\bf Part 3.}

Follows from Line \ref{ln:z_z_w} in the Algorithm \ref{alg:fourier_set_query_I}, we have that
\begin{align*}
   \wh{z}^{(i+1)} = \wh{z}^{(i)} + \wh{w}^{(i)}.
\end{align*}

{\bf Part 4.}

Follows from Line \ref{ln:x_x_sub_z} in the Algorithm \ref{alg:fourier_set_query_I}, we have that
\begin{align*}
   \wh{x}^{(i+1)} = \wh{x} - \wh{z}^{(i+1)}.
\end{align*}

{\bf Part 5.}

By Lemma \ref{lem:fourier_set_query_I_noise_0}, we have that
\begin{align}\label{eq:fourier_set_query_I_noise}
 \Pr \left[ \left\| \wh{x}^{(i)}_{T_i} - \wh{w}^{(i)} \right\|_2^2 \leq  \frac{\epsilon_i}{20} ( \| \wh{x}^{(i)}_{\ov{S}_i} \|_2^2 + \delta^2 n \| \wh{x} \|_1^2 ) \right] \geq 1- \alpha_i.
\end{align}

Recall that 
\begin{align*} 
\wh{w}^{(i)} = \wh{z}^{(i+1)} - \wh{z}^{(i)} = \wh{x}^{(i)} - \wh{x}^{(i+1)}.
\end{align*} 
It is obvious that
\begin{align*}
\supp(\wh{w}^{(i)} ) \subseteq T_i.
\end{align*}

Conditioning on all coordinates in $T_i$ are well isolated and Eq.~\eqref{eq:fourier_set_query_I_noise} holds, we have
\begin{align*}
\| \wh{x}_{\ov{S}_{i+1}}^{(i+1)} \|_2^2 
=& ~ \| ( \wh{x}^{(i)} - \wh{w}^{(i)} )_{\ov{S}_{i+1}} \|_2^2 \\
= & ~ \| \wh{x}^{(i)}_{\ov{S}_{i+1}} - \wh{w}^{(i)}_{\ov{S}_{i+1}} \|_2^2 \\
= & ~ \| \wh{x}^{(i)}_{\ov{S}_{i+1}} - \wh{w}^{(i)} \|_2^2 \\
= & ~ \| \wh{x}^{(i)}_{\ov{S}_{i} \cup T_i} - \wh{w}^{(i)} \|_2^2 \\
= & ~ \| \wh{x}^{(i)}_{\ov{S}_i} \|_2^2 + \| \wh{x}^{(i)}_{T_i} - \wh{w}^{(i)} \|_2^2  \\
\leq & ~ \| \wh{x}^{(i)}_{\ov{S}_i} \|_2^2 + \epsilon_i ( \| \wh{x}^{(i)}_{\ov{S}_i} \|_2^2 + \delta^2 n \| \wh{x} \|_1^2 )  \\
= & ~ (1+\epsilon_i)\| \wh{x}^{(i)}_{\ov{S}_i} \|_2^2 + \epsilon_i \delta^2 n \| \wh{x} \|_1^2.
\end{align*}
where the first step comes from $\hat{x}^{(i+1)}=\hat{x}^{(i)}-\hat{w}^{(i)}$, the second step is due to rearranging the terms, the third step is due to $\hat{w}^{(i)}=\hat{w}^{(i)}_{\ov{S}_{i+1}}$, and the forth step comes from $S_i = T_i \cup S_{i+1}$, the fifth step is due to rearranging the terms, the sixth step the comes from a Eq.~\eqref{eq:fourier_set_query_I_noise}, and the final step comes from merging the $\| \wh{x}^{(i)}_{\ov{S}_i} \|_2^2$ terms.
\end{proof}

\subsection{Induction to all the iterations}
\label{sec:induction}

For completeness, we give the induced result among the all the iterations ($i\in [R]$). By the following lemma at hand, we can finally attain the theorem in Section~\ref{sec:fourier_main_result}. 
\begin{lemma}\label{lem:iterative_loop_analysis_fourier_set_query_1111}
Given parameters $C \geq 1000$, $\gamma \leq 1/1000$. For any $k\geq 1 , \epsilon \in (0,1)$, $R\geq 1$.
For each $i \in [R]$, we define
\begin{align*}
k_i :=&~ k \gamma^{i-1},\\
\epsilon_i := &~ \epsilon (10\gamma)^i, \\
\alpha_i :=&~ 1/(200 i^3) , \\
B_i := &~ C \cdot k_i/ (\alpha_i^2 \epsilon_i ).
\end{align*}
For each $i\in [R]$, we have with probability $1- 10 \alpha_i / \gamma$, we have
\begin{align*}
    | S_{i+1} | \leq k_{i}
\end{align*}
and
\begin{align*}
 \| \wh{x}_{\ov{S}_{i+1}}^{(i+1)} \|_2^2  \leq (1+\epsilon_i) \| \wh{x}_{\ov{S}_{i}}^{(i)} \|_2^2  + \epsilon_i \delta^2 n \| \wh{x} \|_1^2   
\end{align*}
\end{lemma}
\begin{proof}
Our proof can be divided into two parts. At first, we consider the correctness of the inequalities above with $i=1$. And then based on the result we attain above (See Lemma~\ref{lem:iterative_loop_analysis_fourier_set_query_I} ) and inducing over $i\in [n]$, the proof will be complete.
 
By Lemma~\ref{lem:wellisolated}, we have with probability $1- 6 \alpha_1 $, $t$ is well isolated (See Definition~\ref{def:well-isolated}).

{\bf{Part 1.}} 

We have $|S_1|=|S|\leq k=k_i$. (See Definition~\ref{def:fourier_set_query}). And then by Lemma~\ref{lem:fourier_set_query_I_noise_0}, we have that for $i\in [R]$, $|S_{i+1}|\leq k_i$.

{\bf{Part 2.}} 
Given that all coordinates $t\in [n]$  in $T_1$ are well isolated, with probability at least $1-10\alpha_i/\gamma$, we have
\begin{align*}
\| \wh{x}_{\ov{S}_{2}}^{(2)} \|_2^2
= & ~\| ( \wh{x}^{(1)} - \wh{w}^{(1)} )_{\ov{S}_{2}} \|_2^2 \\
= & ~ \| \wh{x}^{(1)}_{\ov{S}_{2}} - \wh{w}^{(1)}_{\ov{S}_{2}} \|_2^2 \\
= & ~ \| \wh{x}^{(1)}_{\ov{S}_{2}} - \wh{w}^{(1)} \|_2^2 \\
= & ~ \| \wh{x}^{(1)}_{\ov{S}_{1} \cup T_1} - \wh{w}^{(1)} \|_2^2 \\
= & ~ \| \wh{x}^{(1)}_{\ov{S}_1} \|_2^2 + \| \wh{x}^{(1)}_{T_1} - \wh{w}^{(1)} \|_2^2  \\
\leq & ~ \| \wh{x}^{(1)}_{\ov{S}_1} \|_2^2 + \epsilon_1 ( \| \wh{x}^{(1)}_{\ov{S}_1} \|_2^2 + \delta^2 n \| \wh{x} \|_1^2 )  \\
= & ~ (1+\epsilon_1)\| \wh{x}^{(1)}_{\ov{S}_1} \|_2^2 + \epsilon_1 \delta^2 n \| \wh{x} \|_1^2.
\end{align*}
where the first step comes from $\hat{x}^{(2)}=\hat{x}^{(1)}-\hat{w}^{(1)}$, the second step is due to rearranging the terms, the third step is due to $\hat{w}^{(1)}=\hat{w}^{(1)}_{\ov{S}_{2}}$, and the forth step comes from $S_1 = T_1 \cup S_{2}$, the fifth step is due to rearranging the terms, the sixth step the comes from expanding the terms, and the final step comes from merging the $\| \wh{x}^{(1)}_{\ov{S}_1} \|_2^2$ terms.

By Lemma~\ref{lem:iterative_loop_analysis_fourier_set_query_I}, for all $i\in [R]$, we can have
\begin{align*}
    \| \wh{x}_{\ov{S}_{i+1}}^{(i+1)} \|_2^2  \leq (1+\epsilon_i) \| \wh{x}_{\ov{S}_{i}}^{(i)} \|_2^2  + \epsilon_i \delta^2 n \| \wh{x} \|_1^2
\end{align*}
\end{proof}

\subsection{Main result}\label{sec:fourier_main_result}

\begin{algorithm*}[!ht]\caption{Fourier set query algorithm}\label{alg:fourier_set_query_I}
\begin{algorithmic}[1]
\Procedure{\textsc{FourierSetQuery}}{$x,S,\epsilon,k$}\Comment{Theorem~\ref{thm:fourier_set_query_I}}
	\State $\gamma \leftarrow 1/1000$, $C \leftarrow 1000$,
	$\wh{z}^{(1)} \leftarrow 0$, $S_1 \leftarrow S$
	\For{ $i=1 \to R$}
		\State $k_i \leftarrow k \gamma^i$, $\epsilon_i \leftarrow \epsilon (10\gamma)^i$, $\alpha_i \leftarrow 1/(100 i^3)$, $B_i \leftarrow C \cdot k_i / (\alpha_i^2 \epsilon_i)$
		\State $\wh{w}^{(i)}, T_i \leftarrow \textsc{EstimateValues}(x,\wh{z}^{(i)},S_i,B_i,\delta,\alpha_i)$ \Comment{$\wh{w}^{(i)}$ is $|T_i|$-sparse}
		\State $S_{i+1} \leftarrow S_i \backslash T_i$
		\State $\wh{z}^{(i+1)} \leftarrow \wh{z}^{(i)} + \wh{w}^{(i)}$ \label{ln:z_z_w}
	\EndFor
	\State \Return $\wh{z}^{(R+1)}$
\EndProcedure
\Procedure{\textsc{EstimateValues}}{$x,\wh{z},S,B,\delta,\alpha$} \Comment{Lemma~\ref{lem:iterative_loop_analysis_fourier_set_query_I}}
		\State Choose $a,b \in [n]$ uniformly at random
		\State Choose $\sigma$ uniformly at random from the set of odd numbers in $[n]$ 
	    \State $\wh{u}\leftarrow \textsc{HashToBins}(P_{\sigma,a,b}, \alpha, \wh{z},B, \delta, x)$
	\State $\wh{w} \leftarrow 0 $, $T\leftarrow \emptyset$
	\For{$t \in S$}
		\If{$t$ is isolated from other coordinates of $S$} \Comment{$h_{\sigma,b}(t) \notin h_{\sigma,b}(S \backslash \{t\})$}
			\If{no large offset} \Comment{$n(1-\alpha)/(2B)>|o_{\sigma,b}(t)|$}
				\State \label{ln:supp_w_S} $\wh{w}_t \leftarrow \wh{u}_{h_{\sigma,b}(t)} e^{-\frac{2\pi \i}{n} \sigma a t }$ \label{alg:wh_w_t_def}
				\State $T \leftarrow T \cup \{t\}$
			\EndIf
		\EndIf
	\EndFor
	\State \Return $\wh{w}, T$
\EndProcedure
\Procedure{\textsc{HashToBins}}{$P_{\sigma,a,b}, \alpha, \wh{z},B, \delta, x$}
	\State Compute $\wh{y}_{jn/B}$ for $j\in [B]$, where $y = G_{B,\alpha,\delta} \cdot (P_{\sigma,a,b} x)$ 
	\State Compute $\wh{y}'_{jn/B} = \wh{y}_{jn/B} - ( \wh{G'_{B,\alpha,\delta} } * \wh{P_{\sigma,a,b} z} )_{jn/B}$ \label{ln:x_x_sub_z}
	\State \Return $\wh{u}_j = \wh{y}'_{jn/B}$
\EndProcedure
\end{algorithmic}
\end{algorithm*}

In this subsection, we give the main result as the following theorem. 

\begin{theorem}[Main result]\label{thm:fourier_set_query_I}
Given a vector $x \in \C^n$ and the $\hat{x}$ as the concrete Fourier transformation result, for every $\epsilon, \delta \in (0,1)$ and $k\geq 1$, any $S \subseteq [n]$, $|S| = k$, there exists an algorithm (Algorithm~\ref{alg:fourier_set_query_I}) that takes 
\begin{align*}
O(\epsilon^{-1} k \log (n/\delta)  )
\end{align*}
samples, runs in 
\begin{align*}
O(\epsilon^{-1} k \log (n/\delta) )
\end{align*}
time, and outputs a vector $x'\in \C^n$ such that
\begin{align*}
\| ( x' - \wh{x} )_S  \|_2^2 \leq \epsilon \| \wh{x}_{\bar{S}} \|_2^2 + \delta \| \wh{x} \|_1^2 
\end{align*}
holds with probability at least $9/10$.
\end{theorem}
\begin{proof}

By the Setting in the Algorithm~\ref{alg:fourier_set_query_I}, we can make the assumption in Lemma~\ref{lem:iterative_loop_analysis_fourier_set_query_I} hold. And by induction on Lemma~\ref{lem:iterative_loop_analysis_fourier_set_query_I}, the following conclusion can be attained by us.

By Lemma~\ref{lem:iterative_loop_analysis_fourier_set_query_I} and the parameters as follows
\begin{align*}
    &k_i := k \gamma^{i-1},\\ &\epsilon_i :=  \epsilon (10\gamma)^i,\\ &\alpha_i = 1/(200 i^3),\\  &B_i := C \cdot k_i/ (\alpha_i^2 \epsilon_i ),
\end{align*}
for $i\in [R]$, we can have that with probability $1- 10 \alpha_i / \gamma$, we have
\begin{enumerate}
    \item $\supp( \wh{w}^{(i)} ) \subseteq S_i$.
    \item $| S_{i+1} | \leq k_{i+1}$.
    \item $\wh{z}^{(i+1)} = \wh{z}^{(i)} + \wh{w}^{(i)}$.
    \item $\wh{x}^{(i+1)} = \wh{x} - \wh{z}^{(i+1)}$.
    \item $\| \wh{x}_{\ov{S}_{i+1}}^{(i+1)} \|_2^2  \leq (1+\epsilon_i) \| \wh{x}_{\ov{S}_{i}}^{(i)} \|_2^2  + \epsilon_i \delta^2 n \| \wh{x} \|_1^2$.
\end{enumerate}

By Lemma~\ref{lem:iterative_loop_analysis_fourier_set_query_1111}, we can conclude that with $R=\log k$ iterations, we will attain the result we want. Then we will give the analysis about the time complexity and sample complexity.

{\bf Proof of Sample Complexity.}

From analysis above, the sample needed in each iteration is $O((B_i / \alpha_i) \log (n/\delta))$ then we have the following complexity.

The sample complexity of \textsc{Estimation} is
\begin{align*} 
\sum_{i=1}^R (B_i / \alpha_i) \log (n/\delta) = O(\epsilon^{-1} k \log (n/\delta)).
\end{align*}
The time in each iteration mainly from two parts. The EstimateValues and HashToBins functions. For the running time of EstimateValues, its running time is mainly from loop. The number of the iterations of the loop can be bounded by $O(B_i/\alpha_i\log(n/\delta))$.

By Lemma~\ref{lem:u_hashbin}, we can attain the time complexity of HashToBins with the bound of $O(B_i/\alpha_i\log(n/\delta))$. This function is used only once at each iteration.

With $R=\log k$, we can have the following equation.

{\bf Proof of Time Complexity.}
The Time complexity of \textsc{Estimation} is
\begin{align*} 
\sum_{i=1}^R (B_i / \alpha_i) \log (n/\delta) = O(\epsilon^{-1} k \log (n/\delta)).
\end{align*}

{\bf Proof of Success Probability.}

The failure probability is $\sum_{i=1}^{R} 10 \alpha_i /\gamma < 1/10.$

{\bf Upper bound $\| \wh{x}_{\ov{S}_{i}}^{(i)} \|_2^2$.} 

By Lemma \ref{lem:iterative_loop_analysis_fourier_set_query_I}, we have that
\begin{align}\label{eq:fourier_set_query_I_upper_bound_ovSi}
\| \wh{x}^{(i)}_{\ov{S}_{i} } \|_2^2 \notag 
\leq & ~ (1+\epsilon_i) \| \wh{x}_{\ov{S}_i}^{(i)} \|_2^2 + \epsilon_i \delta^2 n \| \wh{x} \|_1^2 \notag \\
\leq & ~ (1+\epsilon_i) (1+\epsilon_{i-1}) \| \wh{x}_{\ov{S}_{i-1}}^{(i-1)} \|_2^2 \notag \\
& + ( (1+\epsilon_i) \epsilon_{i-1} + \epsilon_i ) \delta^2 n \| \wh{x} \|_1^2 \notag \\
\leq & ~ \prod_{j=1}^i (1+\epsilon_j) \| \wh{x}_{\ov{S}_j} \|_2^2 + \sum_{j=1}^i \epsilon_j \delta^2 n \| \wh{x} \|_1^2 \prod_{l = j+1}^i (1+\epsilon_l) \notag \\
\leq & ~ 8 ( \| \wh{x}_{\ov{S}_i} \|_2^2 +  \delta^2 n \| \wh{x} \|_1^2 ),
\end{align}

where the first step comes from the assumption in Lemma~\ref{lem:iterative_loop_analysis_fourier_set_query_I}, the second step comes from the assumption in Lemma~\ref{lem:iterative_loop_analysis_fourier_set_query_I}, the third step refers to recursively apply the second step, the last step follows by a geometric sum. 

{\bf Proof of Final Error.}
We can bound the query error by:
\begin{align*}
 \| \wh{x}_S - \wh{z}^{(R+1)} \|_2^2 
= & ~  \sum_{i=1}^R \| \wh{x}_{T_i}^{(i)} - \wh{w}^{(i)} \|_2^2 \\
\leq & ~ \sum_{i=1}^R  k_i \mu^{(i)} /20 \\
\leq & ~ \sum_{i=1}^R \epsilon_i ( \| \wh{x}_{\ov{S}_i}^{(i)} \|_2^2 + \delta^2 n \| \wh{x} \|_1^2 )  /20  \\
\leq & ~ \sum_{i=1}^R \epsilon_i \cdot 10( \| \wh{x}_{\ov{S}} \|_2^2 + \delta^2 n \| \wh{x} \|_1^2 )  /20  \\
\leq & ~ \epsilon (\| \wh{x}_{\ov{S}} \|_2^2 + \delta^2 n \| \wh{x} \|_1^2). 
\end{align*}
where the first step follows that $T_i$ is well isolated (See Definition~\ref{def:well-isolated}.) and $\wh{w}^{(i)} = \wh{z}^{(i+1)} - \wh{z}^{(i)}$, the second step is by Eq.~\eqref{eq:fourier_set_query_I_noise}, the third step comes from the definition of $\mu^{(i)}$ in Eq.~\eqref{eq:mu_def}, the fourth step follows from the Eq.\eqref{eq:fourier_set_query_I_upper_bound_ovSi}, and the final step follows from the geometric sum, $\epsilon_i =  \epsilon (10\gamma)^i$ and $\gamma \leq 1/1000$.
 
\end{proof}

\section{Conclusion}\label{sec:conclusion}

Fourier transformation is an intensively researched topic in a variety of scientific disciplines. Numerous applications exist within machine learning, signal processing, compressed sensing, etc.
In this paper, we study the problem of Fourier set query. With an approximation parameter $\epsilon$, a vector $x \in \mathbb{C}^{n}$ and a query set $S \subset [n]$ of size $k$, our algorithm uses $O(\epsilon^{-1} k \log(n/\delta))$ Fourier measurements, runs in $O(\epsilon^{-1} k \log(n/\delta))$  time and outputs a vector $x'$ such that $\| ( x' - \wh{x} )_S  \|_2^2 \leq \epsilon \| \wh{x}_{\bar{S}} \|_2^2 + \delta \| \wh{x} \|_1^2 $ with probability of at least $9/10$.

\ifdefined\isarxivversion
\bibliographystyle{alpha}
\bibliography{ref}
\else
\bibliography{ref}
 
\bibliographystyle{IEEEtran}

\fi





\end{document}